\documentclass[twocolumn, switch]{article} 

\usepackage{preprint}

\usepackage{amsmath, amsthm, amssymb, amsfonts}

\usepackage[numbers,square]{natbib}
\usepackage[utf8]{inputenc}	
\usepackage[T1]{fontenc}	
\usepackage{xcolor}		
\usepackage[colorlinks = true,
            linkcolor = purple,
            urlcolor  = blue,
            citecolor = cyan,
            anchorcolor = black]{hyperref}	
\usepackage{booktabs} 		
\usepackage{nicefrac}		
\usepackage{microtype}		
\usepackage{lineno}		
\usepackage{float}			
\usepackage{multicol}		

\usepackage{lipsum}		

\usepackage{newfloat}
\DeclareFloatingEnvironment[name={Supplementary Figure}]{suppfigure}
\usepackage{sidecap}
\sidecaptionvpos{figure}{c}

\usepackage{titlesec}
\titlespacing\section{0pt}{12pt plus 3pt minus 3pt}{1pt plus 1pt minus 1pt}
\titlespacing\subsection{0pt}{10pt plus 3pt minus 3pt}{1pt plus 1pt minus 1pt}
\titlespacing\subsubsection{0pt}{8pt plus 3pt minus 3pt}{1pt plus 1pt minus 1pt}

\usepackage{tikz,xcolor,hyperref}

\definecolor{lime}{HTML}{A6CE39}
\DeclareRobustCommand{\orcidicon}{
	\begin{tikzpicture}
	\draw[lime, fill=lime] (0,0) 
	circle [radius=0.16] 
	node[white] {{\fontfamily{qag}\selectfont \tiny ID}};
	\draw[white, fill=white] (-0.0625,0.095) 
	circle [radius=0.007];
	\end{tikzpicture}
	\hspace{-2mm}
}
\foreach \x in {A, ..., Z}{\expandafter\xdef\csname orcid\x\endcsname{\noexpand\href{https://orcid.org/\csname orcidauthor\x\endcsname}
			{\noexpand\orcidicon}}
}

\title{Automated Cache for Container Executables}

\usepackage{xwatermark}
\newwatermark[firstpage,color=gray!60,angle=90,scale=0.32, xpos=-4.05in,ypos=0]{\href{https://doi.org/}{\color{gray}{}}}
\newwatermark[firstpage,color=gray!60,angle=90,scale=0.32, xpos=3.9in,ypos=0]{\href{https://doi.org/}{\color{gray}{Preprint doi}}}
\newwatermark[firstpage,color=gray!90,angle=0,scale=0.28, xpos=0in,ypos=-5in]{*correspondence: \texttt{sochat1@llnl.gov}}

\usepackage{authblk}

\author[1]{Vanessa Sochat\orcidA{}}
\author[2]{Matthieu Muffato\orcidB{}}
\author[3]{Audrey Stott\orcidC{}}
\author[3]{Marco {De La Pierre\orcidD{}}}
\author[4]{Georgia Stuart\orcidE{}}

\affil[1]{Computer Scientist, Lawrence Livermore National Lab, USA}
\affil[2]{Tree of Life, Wellcome Sanger Institute, Cambridge, UK}
\affil[3]{Pawsey Supercomputing Research Institute, Kensington, WA, Australia}
\affil[4]{UT Dallas, Richardson, TX, USA}

\usepackage{listings}
\usepackage{color}

\definecolor{dkgreen}{rgb}{0,0.6,0}
\definecolor{gray}{rgb}{0.5,0.5,0.5}
\definecolor{mauve}{rgb}{0.58,0,0.82}

\begin{document}

\twocolumn[ 
  \begin{@twocolumnfalse} 
  
\maketitle

\begin{abstract}
Linux container technologies such as Docker \cite{docker} and Singularity \cite{singularity} offer encapsulated environments for easy execution of software. In high performance computing, this is especially important for evolving and complex software stacks with conflicting dependencies that must co-exist. Singularity Registry HPC (``shpc'') \cite{shpc} was created as an effort to install containers in this environment as modules \cite{modules}, seamlessly allowing for typically hidden executables inside containers to be presented to the user as commands, and as such significantly simplifying the user experience. A remaining challenge, however, is deriving the list of important executables in the container. In this work, we present new automation and methods that allow for not only discovering new containers in large community sets, but also deriving container entries with important executables. With this work we have added over 8,000 containers from the BioContainers \cite{biocontainers} community that can be maintained and updated by the software automation over time. All software is publicly available on the GitHub platform, and can be beneficial to container registries and infrastructure providers for automatically generating container modules to lower the usage entry barrier and improve user experience.
\end{abstract}


\vspace{0.35cm}

  \end{@twocolumnfalse} 
] 

\section*{(1) Overview}
\label{sec:overview}

\section{Introduction}
\label{sec:introduction}

Containerization technologies \cite{docker,singularity} have been a game-changer for reproducible research, allowing users to not only build and use reproducible environments on high performance computing (HPC) systems, but also allowing for saving the container binaries in a registry for others in the future to do the same. As HPC has matured over the decades, tools have been created that make it easier to install software to this environment, whether that be in the form of a package manager \cite{spack,easybuild} or module software \cite{modules,lmod}, which allow a user to load a namespace of commands to use. As most package managers and modules typically install software from source, the Singularity Registry HPC (shpc) \cite{shpc} software was created as a marriage between the two - allowing HPC administrators and users to install containers as modules.

\subsection{The Design of Singularity Registry HPC}

Singularity Registry HPC provides a manager for registry entries that can seamlessly install to module software and then render itself no longer needed. At its initial creation, each registry entry was only allowed to be a YAML file \cite{yaml} in a local filesystem folder organized by the container unique resource identifier namespace. An example with the unique resource identifier ``quay.io/biocontainers/samtools'' is shown below.

\begin{lstlisting}
$ tree quay.io/biocontainers/samtools/
quay.io/biocontainers/samtools/
> container.yaml
\end{lstlisting}

The entry provides a main ``container.yaml'' file with an installation source (e.g., a container registry), a maintainer and description, along with container tags and digests \cite{tag-vs-digest}, and aliases or named paths \cite{shpc}. An example alias - a mapping of a term ``samtools'' to a path is shown below:

\begin{lstlisting}
aliases:
  samtools: /usr/local/bin/samtools
\end{lstlisting}
  
This named path makes it easy to find useful executables within a container. The shpc software is able to provide a set of aliases that are made available to users as shell commands. This significantly lowers the barrier to container adoption in HPC, as users need to know almost nothing about container usage and syntax. For the example above, loading the container as a module would expose multiple aliases, the main one of interest likely the ``samtools'' alias for the user to interact with. An example of the underlying command that the user would need to know without the shpc software is shown below using the Singularity \cite{singularity} container technology.

\begin{lstlisting}
$ singularity <singularity-options> exec <options> -B <bind> <container> /usr/local/bin/samtools "$@"
\end{lstlisting}

The shpc software provides equivalent aliases for singularity shell, run and various inspection commands, and also supports custom container options and binds. The original set of approximately 200 entries was derived manually by the creator and contributors, and recognizing the constant change of container tags, a GitHub action was quickly developed to retrieve updated tags via the container registry API endpoint \cite{shpc}. As the design evolved, there was desire for more automation and separation of responsibility between the manager software and the registry entries. As a result, the software was updated to support local or remote registries, and the main registry, along with the monthly automation to update tags, moved into a remote version-controlled repository \cite{shpc-registry}. 

This decoupling of the manager software from the underlying registry presented a new opportunity to extend automation. Previously, updating the entries of a registry required pulling the updated software along with the container entries from the version-control source. With a remote endpoint that serves a static application programming interface (API), this is no longer required -- the user's local registry can retrieve the latest registry entries without updating the software, and the entries can be updated separately with tagged releases provided monthly. However, while updating tags in registry entries was automated, creating an entirely new registry entry was not. This presented a new challenge for automation, not only for the manual addition of a new container entry requested by a user, but also for the addition of potentially thousands of containers from an external source. Complete automation would not only require deriving container tags and digests, but also the executable paths.  In this work, we walk through the steps and methodology used to first semi\-automate addition of a single entry, and then to automate adding thousands of containers from the BioContainers set \cite{biocontainers}. The newly added 8,000+ containers are available for install using the shpc software, and all work is publicly available with active automation to keep the registry updated.

\section{Implementation and architecture}
\label{sec:methods}

\subsection{Automated Recipe Generation}

A semi-automated solution to generate a container entry to the online registry was introduced \cite{shpc} as a workflow. It takes a container unique resource identifier, description, and website reference, and generates a pull request to the repository to add the container. To support this workflow, we developed the ``guts'' software \cite{guts} that knows how to pull a container, retrieve and parse the ``PATH'' from the image manifest \cite{oci-manifest}, and then dump the container filesystem to a temporary location to search those locations statically and clean up. This procedure makes the assumption that the developer of the container has added locations of important executables to the ``PATH'' variable. An additional filtering step is needed to discover important executables (and not those found in the operating system provided by the container). We do an additional ``diff'' \cite{diff} with known executables from the most common base containers, which are provided via another automated workflow that is updated nightly \cite{shpc-guts}. The final set of executables represents a set that is unique to the changes developers made from the base containers.

To retrieve updated tags, we make a request to the Docker API \cite{dockerapi} to retrieve a list of image tags, and retrieve associated digests via image manifests \cite{oci-manifest}. The ``pipelib'' software \cite{pipelib} is then used to sort tags by semantic versions, allowing us to filter tags and identify the ``latest,'' a special tag provided by the shpc software as the default version to install. With this workflow to automatically derive tags and executables given any container identifier, it became possible to request a recipe for a specific container directly on the remote registry repository, and then receive a pull request with a prepared registry entry. Without any further filtering of the executables (e.g., installed dependencies unique to the container that are not of interest), the entry typically requires additional curation to filter down the discovered executables to those that are the most important. This semi-automated workflow allows for easy addition of hand-picked containers, but would be substantial work to add tens to thousands more.

\subsection{Container Executable Frequency}

With a request to make available more BioContainers \cite{biocontainers}, a next logical step was to figure out how to combine the semi-automated generation with a means not only to add an individual container, but also to add potentially thousands of containers from an external source. At this scale, manual edits could not be required -- it would need to be possible to identify the most important executables without human curation. Addressing this challenge would require better understanding of the distribution of executables across containers, and then determining a strategy to identify the most unique to a container. Toward this aim, we first developed a cache -- the Singularity Registry HPC cache -- to store a complete list of executables on the ``PATH'' across all BioContainers \cite{shpc-registry-cache}. The cache is enabled by a generalized container binary discovery workflow designed to work on GitHub \cite{container-discovery-action}, called an action. First, the action is provided with an updated list of container binaries provided by the Galaxy Project \cite{afgan2018galaxy}. For each container identifier, we then again use the guts software to discover all binaries on the ``PATH,'' and save a JSON data file of the binaries to the repository, organized again by the container identifier. As a final step, we derive a second JSON data file with counts of executable names across containers. This counts data file can next be used alongside the shpc remote registry to intelligently filter down an entire set of executables in one container to a more relevant set. This entire set of actions is enabled by a few lines added to a workflow file, as shown in the shpc remote registry workflow \cite{shpc-registry-workflow}.

\subsection{Automated Scaled Recipe Generation}

Given the availability of summary counts for over 8,000 BioContainers, we can deploy the following algorithm to generate a list of meaningful executables per container. This assumes that we have a regularly updated executable count cache derived from the Galaxy Project listing.

\begin{enumerate}
\item{Identify a new container, C, not in the registry from the executable cache}
\item{Create a set of global executable counts, G}
\item{Define a set of counts from G in C as S}
\item{Rank order S from least to greatest}
\item{Include any entries in S that have a frequency < 10}
\item{Include any entries in S that have any portion of the name matching the container identifier}
\item{Above that, add the next 25 executables with the lowest frequencies, and < 1,000}
\end{enumerate}

The algorithm above assumes that the most unique executables in a container are less likely to appear in other containers, represented by a lower frequency. Always including executables that appear fewer than 10 times across the entire dataset allows for a container to have many unique commands. We chose these thresholds based on manual testing and visualization of the final list of executables, and found that these steps produced the set of binaries that we would expect or want for manual curation. We can combine programmatically derived tags and digests with these container aliases and other automated metadata to generate a final ``container.yaml''. From this YAML file, the shpc software can install the module to an HPC system and generate the respective executables as module commands.

\subsection{Automated Recipe Updates}

The original workflow to automatically update container tags and digests uses a native ``update'' command provided by the Singularity Registry HPC client, and this was run once a month across all containers in the current registry directly before a monthly release. However, with the addition of 8,000+ containers this monthly update would no longer be feasible within the 6 hour limit of a GitHub action runner \cite{runner}. To address this challenge, we developed a simple strategy to break up an entire list of container identifiers into equal groups, and have those groups remain consistent even given new additions to the registry. To do this, we first generate hashes for each of our container identifiers, and then generate hexdigests \cite{hashlib} that we convert into integer numbers. Then we take the modulus of the minimum number of days that can possibly appear in any month (N=28) to assign each number into a specific group in the range 0-27. We add 1 to this number for a range that matches with days of the month, 1-28.  On a high level, this means that we can reliably split our container identifiers into equal groups, each of which is matched to a specific day of the month. In our workflow, we can then derive the groups, take the subset for the day the workflow is running, and update that set.  This algorithm is represented and provided in a GitHub action \cite{split-list-action} for the interested reader.

\section{Applications}
\label{sec:applications}


We took this work to the Pawsey Supercomputing Research Centre (Pawsey) \cite{pawsey}, a tier-1 Australian national high performance computing facility, where having these BioContainers made available as modules is perceived to vastly improve the accessibility and usage of containers in the life sciences. Through their involvement in the Australian BioCommons \cite{biocommons} Bring-Your-Own-Device Expansion Project, earlier phase discussions and surveys have highlighted repeatedly that containers are an integral part of life science research, but uptake is impeded by the lack of knowledge, confusion and time in learning about containers. Pawsey role was to provide technical and compute expertise for user access to the Galaxy Project’s repository of Singularity images of BioContainers through a read-only filesystem called CernVM-FS \cite{cvmfs}. While this filesystem cache significantly reduces duplication of images and time for building, researchers still face the hurdle of container syntax. 

Our automation of shpc recipes for BioContainers means that Pawsey, as well as other tier-1 and tier-2 partners of the Australian BioCommons, can easily install and have the same list of over 8,000 BioContainers loaded as shpc modules. With a simple script to match a discovered container in the filesystem to an shpc container entry \cite{biocontainers-match-script}, these compute facilities can utilize the existing library of Singularity images through their CernVM-FS filesystem. Recipe updates provided by shpc also ensure that all new versions of BioContainers, while being added to the CernVM-FS repository, are simultaneously made available to their researchers as shpc modules.

\section{Quality control}
\label{sec:quality}

For Singularity Registry HPC, tests are run via continuous integration for each pull request into the main branch by means of GitHub Actions.  Tests span all functionality of the software across several versions of module software and container technologies. 

The container registry updates and additions are done via an automated workflow, and manually checked by the main developer, author VS, for any changes. Lists of executables provided in the cache are spot checked by developers to ensure what is expected is there (e.g., a samtools container should minimally have the executable for samtools). Feedback comes in from the user base about executables that might be removed or added to further tweak added container recipes.

\section*{(2) Availability}
\label{sec:availability}
\setcounter{section}{0}

\section*{Operating system} 
Singularity Registry HPC and associated tooling should work on most Unix and Linux flavored distributions. The software was developed on Ubuntu 22.04.

\section*{Programming language}
This set of tools is developed to support Python 3.7 and higher. Python 2.x is not supported.

\section*{Dependencies}
The newly released cache and automation can run on GitHub actions with the environment encapsulated by the runner.
The shpc set of tools requires the requests library, jsonschema, and generally expects module software to be installed \cite{lmod, modules}. See the ``version.py'' in each project for details. Naturally, shpc also requires a container execution runtime, such as Singularity  \cite{singularity}, Podman \cite{podman} or Docker \cite{docker}.

\section*{List of contributors}
Authors VS, MM, AS, MDLP, and GS have all contributed directly to the software.

\section*{Software location: Archive}
\begin{itemize}
    \item Name: singularity-hpc
    \begin{itemize}
        \item Persistent Identifier:\newline \href{https://github.com/singularityhub/singularity-hpc/archive/refs/tags/0.1.16.tar.gz}{singularityhub/singularity-hpc}
        \item License: MPL-2.0
        \item Publisher: Vanessa Sochat
        \item Version published: 0.1.16
        \item Date published: 5 November 2022
    \end{itemize}    
\end{itemize}
\begin{itemize}
    \item Name: shpc-registry
    \begin{itemize}
        \item Persistent Identifier: \newline \href{https://github.com/singularityhub/shpc-registry/archive/refs/tags/2022-11.tar.gz}{singularityhub/shpc-registry}
        \item License: MPL-2.0
        \item Publisher: Vanessa Sochat
        \item Version published: 2022-11
        \item Date published: 2 November 2022
    \end{itemize}    
\end{itemize}
\begin{itemize}
    \item Name: shpc-registry-cache
    \begin{itemize}
        \item Persistent Identifier: \newline \href{https://github.com/singularityhub/shpc-registry-cache/archive/refs/tags/2022-10.tar.gz}{singularityhub/singularity-hpc}
        \item License: MPL-2.0
        \item Publisher: Vanessa Sochat
        \item Version published: 2022-10
        \item Date published: 21 November 2022
    \end{itemize}    
\end{itemize}

\section*{Software location: Code repository}
\begin{itemize}
    \item Name: singularity-hpc
    \begin{itemize}
        \item Persistent Identifier: \newline \href{https://github.com/singularityhub/singularity-hpc}{singularityhub/singularity-hpc}
        \item License: MPL-2.0
        \item Date published: 3 April 2021
    \end{itemize}    
\end{itemize}
\begin{itemize}
    \item Name: shpc-registry
    \begin{itemize}
        \item Persistent Identifier: \newline \href{https://github.com/singularityhub/shpc-registry}{singularityhub/shpc-registry}
        \item License: MPL-2.0
        \item Date published: 31 July 2022
    \end{itemize}    
\end{itemize}
\begin{itemize}
    \item Name: shpc-registry-cache
    \begin{itemize}
        \item Persistent Identifier: \newline \href{https://github.com/singularityhub/shpc-registry-cache}{singularityhub/shpc-registry-cache}
        \item License: MPL-2.0
        \item Date published: 18 October 2022
    \end{itemize}    
\end{itemize}

\section*{Software location: Language}
English.

\section*{(3) Reuse potential}
\label{sec:reuse}
\setcounter{section}{0}

The documentation pages \cite{shpc-docs} for Singularity Registry HPC have been updated to include a developer tutorial \cite{shpc-dev-tutorial} that guides through the steps to setup and deploy a registry of container modules based on automatic recipe generation and executable lookup. Separate automated workflows exist for cache generation \cite{container-discovery-cache-creation-action} and for using a cache to update a registry \cite{shpc-registry-cache, container-discovery-action}. In this way, container registries and infrastructures providers can reuse the software presented in this manuscript to automatically generate a registry for a custom collection of containers, which is of relevance for their operations and activities. Some parameters of the protocol can be modified to tune the resulting registry; for instance, this flexibility applies to the filtering of container tags/digest, and to the selection of container executable based on executable frequencies.

Two package modules that constitute part of the presented toolkit, pipelib and guts, can be reused independently for purposes that go beyond the generation of container registries. The former offers functionalities to filter containers tags and digests, whereas the latter allows to extract information on available executables in a container.

Finally, it is also worth mentioning the usability impact of the real use case presented as application for the presented software. The collection of 8,000 recipes to generate container modules for the whole collection of BioContainers represents a valuable, impactful resource for the community of life scientists, as well as that of infrastructure providers and support staff that offer services to the former.

All packages are open source on GitHub, and contributions and ideas are welcome.

\section*{(4) Conclusion}
\label{sec:conclusion}
\setcounter{section}{4}

\subsection{Summary}

In this work, we present complete automation to support and continually update a set of over 8,000 containers to install to an HPC system using the Singularity Registry HPC software. Our interesting contributions that we desire to share with the community include:

\begin{itemize}
\item{Singularity Registry HPC, with support for remote registries and automated updates \cite{shpc}}
\item{A self-updating, version-controlled static container registry and API of container metadata \cite{shpc-registry}}
\item{A self-updating, version-controlled database of executable frequencies \cite{shpc-registry-cache}}
\item{The guts software to extract container executables on the ``PATH'' \cite{guts}}
\item{The pipelib software to intelligently filter and sort container tags \cite{pipelib}}
\item{A library of over 8,000 containers to install to an HPC system with Singularity Registry HPC \cite{shpc}}
\end{itemize}

This manuscript presents as a strong example of a research software paper, as the primary focus is on the development of workflows, interfaces, and software to support installing software to complex environments. We hope that any of the automation, data, or software presented is of use or interest to the larger community.

\subsection{Discussion}

The algorithm presented in this paper is a necessity caused by the lack of standard metadata for describing the content of a package or container. Ideally, there should be a machine-readable manifest that would list the primary content (installed by ``make install'' or equivalent), the dependencies, and the base image.
This could be tackled first within the Conda \cite{conda} system. The Conda build system knows which binaries are installed by a given package and what its dependencies are. The listing could be exposed at the package level. Such metadata would take research software closer to the ``FAIR principles for research software'' \cite{fair4rs} (FAIR stands for Findable, Accessible, Interoperable, Reusable). All BioConda packages (and many from other channels) are automatically turned into Docker images by automation at BioContainers. The build could load those manifests into standard container labels that shpc could then use to derive the commands to expose.

It is expected that some recipes created by the algorithm have too few or too many aliases. Being the official shpc registry openly hosted on GitHub, contributions are welcome in the form of pull requests to modify the list of aliases, and we invite the research community to report any error they find. This curation process will happen concurrently to the regular update of tags and digests.

\section*{Contributions}

Author VS designed and created the Singularity Registry HPC software, pipelib, and guts, and the automation detailed in this work. MM, AS, MDLP, and GS contributed to the codebase and design of the Singularity Registry HPC software and the automation.

\section*{Acknowledgements}

We are grateful to the larger HPC community for bug reports, feature requests, and helping to strengthen our community! As this resource has grown out of community need, the authors encourage interaction via issues or discussion on the respective repository, and requests for additions or features.

MM is funded by the Wellcome Trust Grant 218328. AS and MDLP acknowledge support from the Australian BioCommons BYOD Expansion Project, which is funded through NCRIS investments from Bioplatforms Australia and the Australian Research Data Commons (https://doi.org/10.47486/PL105).

\bibliographystyle{ieeetr}

\bibliography{references}  

\begin{thebibliography}{34}
\providecommand{\natexlab}[1]{#1}
\providecommand{\url}[1]{\texttt{#1}}
\expandafter\ifx\csname urlstyle\endcsname\relax
  \providecommand{\doi}[1]{doi: #1}\else
  \providecommand{\doi}{doi: \begingroup \urlstyle{rm}\Url}\fi

\bibitem[Ratliff(2022)]{docker}
James Ratliff.
\newblock {Docker: Accelerated, Containerized Application Development}.
\newblock \url{https://www.docker.com/}, 2022.
\newblock Accessed: 2022-10-25.

\bibitem[Kurtzer et~al.(2017)Kurtzer, Sochat, and Bauer]{singularity}
Gregory~M Kurtzer, Vanessa Sochat, and Michael~W Bauer.
\newblock {Singularity: Scientific containers for mobility of compute}.
\newblock \emph{PLoS One}, 12\penalty0 (5):\penalty0 e0177459, May 2017.

\bibitem[Sochat and Scott(2021)]{shpc}
Vanessa Sochat and Alec Scott.
\newblock {Collaborative Container Modules with Singularity Registry HPC}.
\newblock \emph{Journal of Open Source Software}, 6\penalty0 (63):\penalty0
  3311, 2021.
\newblock \doi{10.21105/joss.03311}.
\newblock URL \url{https://doi.org/10.21105/joss.03311}.

\bibitem[Furlani(1991)]{modules}
John~L Furlani.
\newblock {Modules: Providing a flexible user environment}.
\newblock In \emph{Proceedings of the fifth large installation systems
  administration conference (LISA V)}, pages 141--152, 1991.

\bibitem[da~Veiga~Leprevost et~al.(2017)da~Veiga~Leprevost, Gr{\"u}ning,
  Alves~Aflitos, R{\"o}st, Uszkoreit, Barsnes, Vaudel, Moreno, Gatto, Weber,
  Bai, Jimenez, Sachsenberg, Pfeuffer, Vera~Alvarez, Griss, Nesvizhskii, and
  Perez-Riverol]{biocontainers}
Felipe da~Veiga~Leprevost, Bj{\"o}rn~A Gr{\"u}ning, Saulo Alves~Aflitos,
  Hannes~L R{\"o}st, Julian Uszkoreit, Harald Barsnes, Marc Vaudel, Pablo
  Moreno, Laurent Gatto, Jonas Weber, Mingze Bai, Rafael~C Jimenez, Timo
  Sachsenberg, Julianus Pfeuffer, Roberto Vera~Alvarez, Johannes Griss,
  Alexey~I Nesvizhskii, and Yasset Perez-Riverol.
\newblock {BioContainers: an open-source and community-driven framework for
  software standardization}.
\newblock \emph{Bioinformatics}, 33\penalty0 (16):\penalty0 2580--2582, August
  2017.

\bibitem[Gamblin et~al.(2015)Gamblin, LeGendre, Collette, Lee, Moody,
  De~Supinski, and Futral]{spack}
Todd Gamblin, Matthew LeGendre, Michael~R Collette, Gregory~L Lee, Adam Moody,
  Bronis~R De~Supinski, and Scott Futral.
\newblock {The Spack package manager: bringing order to HPC software chaos}.
\newblock In \emph{Proceedings of the International Conference for High
  Performance Computing, Networking, Storage and Analysis}, pages 1--12, 2015.

\bibitem[Hoste et~al.(2012)Hoste, Timmerman, Georges, and De~Weirdt]{easybuild}
Kenneth Hoste, Jens Timmerman, Andy Georges, and Stijn De~Weirdt.
\newblock {Easybuild: Building software with ease}.
\newblock In \emph{2012 SC Companion: High Performance Computing, Networking
  Storage and Analysis}, pages 572--582. IEEE, 2012.

\bibitem[lmo(2022)]{lmod}
{Lmod: A New Environment Module System --- Lmod 8.7.13 documentation}.
\newblock \url{https://lmod.readthedocs.io/en/latest/}, 2022.
\newblock Accessed: 2022-10-25.

\bibitem[yam(2022)]{yaml}
{YAML Specification Index}.
\newblock \url{https://yaml.org/spec/}, 2022.
\newblock Accessed: 2022-10-25.

\bibitem[Savanth(2021)]{tag-vs-digest}
Vijay Savanth.
\newblock Docker images: Name vs. tag vs. digest.
\newblock \url{https://hackernoon.com/docker-images-name-vs-tag-vs-digest},
  September 2021.
\newblock Accessed: 2022-11-4.

\bibitem[shp(2022{\natexlab{a}})]{shpc-registry}
{shpc-registry: Testing a remote registry for Singularity Registry HPC}.
\newblock \url{https://github.com/singularityhub/shpc-registry},
  2022{\natexlab{a}}.

\bibitem[gut(2022)]{guts}
{guts: Actions and client to derive container guts!}
\newblock \url{https://github.com/singularityhub/guts}, 2022.

\bibitem[oci(2022)]{oci-manifest}
{opencontainers/image-spec manifest}.
\newblock
  \url{https://github.com/opencontainers/image-spec/blob/main/manifest.md},
  2022.

\bibitem[{Wikipedia contributors}(2022)]{diff}
{Wikipedia contributors}.
\newblock diff.
\newblock
  \url{https://en.wikipedia.org/w/index.php?title=Diff&oldid=1118832554},
  October 2022.
\newblock Accessed: NA-NA-NA.

\bibitem[shp(2022{\natexlab{b}})]{shpc-guts}
{shpc-guts: Singularity Registry HPC... container guts!}
\newblock \url{https://github.com/singularityhub/shpc-guts},
  2022{\natexlab{b}}.

\bibitem[doc(2020)]{dockerapi}
Docker hub {HTTP} {API} {V2}.
\newblock \url{https://docs.docker.com/registry/spec/api/}, November 2020.
\newblock Accessed: 2020-11-23.

\bibitem[pip(2022)]{pipelib}
{pipelib: a library for creating pipelines for parsing, filtering, and sorting
  iterables.}
\newblock \url{https://github.com/vsoch/pipelib}, 2022.
\newblock Accessed: 2022-10-26.

\bibitem[shp(2022{\natexlab{c}})]{shpc-registry-cache}
{shpc-registry-cache: A cache of commands (currently for biocontainers)}.
\newblock \url{https://github.com/singularityhub/shpc-registry-cache},
  2022{\natexlab{c}}.

\bibitem[con(2022{\natexlab{a}})]{container-discovery-action}
{container-executable-discovery action}.
\newblock
  \url{https://github.com/singularityhub/container-executable-discovery/},
  2022{\natexlab{a}}.
\newblock Accessed: 2022-11-29.

\bibitem[Afgan et~al.(2018)Afgan, Baker, Batut, Van Den~Beek, Bouvier,
  {\v{C}}ech, Chilton, Clements, Coraor, Gr{\"u}ning, et~al.]{afgan2018galaxy}
Enis Afgan, Dannon Baker, B{\'e}r{\'e}nice Batut, Marius Van Den~Beek, Dave
  Bouvier, Martin {\v{C}}ech, John Chilton, Dave Clements, Nate Coraor,
  Bj{\"o}rn~A Gr{\"u}ning, et~al.
\newblock {The Galaxy platform for accessible, reproducible and collaborative
  biomedical analyses: 2018 update}.
\newblock \emph{Nucleic acids research}, 46\penalty0 (W1):\penalty0 W537--W544,
  2018.

\bibitem[shp(2022{\natexlab{d}})]{shpc-registry-workflow}
{shpc-registry workflow to update Biocontainers}.
\newblock
  \url{https://github.com/singularityhub/shpc-registry/blob/main/.github/workflows/update-biocontainers.yaml},
  2022{\natexlab{d}}.
\newblock Accessed: 2022-11-29.

\bibitem[run(2022)]{runner}
{Usage limits, billing, and administration}.
\newblock
  \url{https://docs.github.com/en/actions/learn-github-actions/usage-limits-billing-and-administration},
  2022.

\bibitem[has(2022)]{hashlib}
hashlib --- secure hashes and message digests --- python 3.11.0 documentation.
\newblock \url{https://docs.python.org/3/library/hashlib.html}, 2022.
\newblock Accessed: 2022-11-4.

\bibitem[{Vanessasaurus}(2022)]{split-list-action}
{Vanessasaurus}.
\newblock {split-list-action: Simple {GitHub} action to evenly split (and
  present a subset) of items based on random selection or the day of the
  month!}
\newblock \url{https://github.com/vsoch/split-list-action}, 2022.

\bibitem[paw(2022)]{pawsey}
{Pawsey Supercomputing Research Centre}.
\newblock \url{https://pawsey.org.au}, 2022.
\newblock Accessed: 2022-11-8.

\bibitem[bio(2022{\natexlab{a}})]{biocommons}
{Australian BioCommons}.
\newblock \url{https://biocommons.org.au}, 2022{\natexlab{a}}.
\newblock Accessed: 2022-11-8.

\bibitem[cvm(2022)]{cvmfs}
{CernVM File System}.
\newblock \url{https://cernvm.cern.ch/fs/}, 2022.
\newblock Accessed: 2022-11-8.

\bibitem[bio(2022{\natexlab{b}})]{biocontainers-match-script}
biocontainer-match.py at main shpc repository.
\newblock
  \url{https://github.com/singularityhub/singularity-hpc/blob/main/example/biocontainer-match.py},
  2022{\natexlab{b}}.
\newblock Accessed: 2022-11-16.

\bibitem[Walsh(2022)]{podman}
Daniel Walsh.
\newblock {The Pod Manager tool (podman)}.
\newblock \url{https://podman.io/}, 2022.
\newblock Accessed: 2022-11-26.

\bibitem[shp(2022{\natexlab{e}})]{shpc-docs}
Singularity registry hpc -- documentation.
\newblock \url{https://singularity-hpc.readthedocs.io}, 2022{\natexlab{e}}.
\newblock Accessed: 2022-12-8.

\bibitem[shp(2022{\natexlab{f}})]{shpc-dev-tutorial}
Singularity registry hpc -- developer tutorial.
\newblock
  \url{https://singularity-hpc.readthedocs.io/en/latest/getting_started/developer-guide.html#developer-tutorial},
  2022{\natexlab{f}}.
\newblock Accessed: 2022-12-14.

\bibitem[con(2022{\natexlab{b}})]{container-discovery-cache-creation-action}
{container-executable-discovery cache creation action}.
\newblock
  \url{https://github.com/singularityhub/shpc-registry-cache/blob/main/.github/workflows/update-cache.yaml},
  2022{\natexlab{b}}.
\newblock Accessed: 2022-11-29.

\bibitem[con(2022{\natexlab{c}})]{conda}
{Conda: Package, dependency and environment management for any language}.
\newblock \url{https://docs.conda.io/}, 2022{\natexlab{c}}.

\bibitem[Barker et~al.(2022)Barker, Chue~Hong, Katz, Lamprecht, Martinez-Ortiz,
  Psomopoulos, Harrow, Castro, Gruenpeter, Martinez, and Honeyman]{fair4rs}
Michelle Barker, Neil~P. Chue~Hong, Daniel~S. Katz, Anna-Lena Lamprecht, Carlos
  Martinez-Ortiz, Fotis Psomopoulos, Jennifer Harrow, Leyla~Jael Castro, Morane
  Gruenpeter, Paula~Andrea Martinez, and Tom Honeyman.
\newblock {Introducing the FAIR Principles for research software}.
\newblock \emph{Scientific Data}, 9:\penalty0 622, October 2022.

\end{thebibliography}






\end{document}